\documentstyle[aps,prl,epsfig,twocolumn]{revtex}
\newcommand{\be}{\begin{equation}}
\newcommand{\ee}{\end{equation}}
\newcommand{\br}{\begin{eqnarray}}
\newcommand{\er}{\end{eqnarray}}
\newcommand{\la}{\langle}
\newcommand{\ra}{\rangle}
\begin{document}
\twocolumn[\hsize\textwidth\columnwidth\hsize\csname
  @twocolumnfalse\endcsname
\title{An alternate model for protective measurements of two-level systems.}
\author{Anirban Das\dag} 
\address{Indian Institute of Technology, Kharagpur, India}
\author{N.D. Hari Dass\ddag} 
\address{Institute of Mathematical Sciences, Chennai, India}
\maketitle
\begin{abstract} 
In this article we propose an alternate model for the so called {\it protective measurements}, more appropriately {\it adiabatic measurements}   
of a spin $\frac{1}{2}$ system where the {\it apparatus} is also a quantum system 
with a {\em finite dimensional Hilbert space}. This circumvents several technical
as well as conceptual issues that arise when dealing with an infinite dimensional
Hilbert space as in the analysis of conventional Stern-Gerlach experiment. Here also it is demonstrated that the response of the detector is 
continuous and it {\it directly} measures {\em expectation values without 
altering the state of the system}(when the unknown original state is a 
{\it nondegenerate eigenstate of the system Hamiltonian}, in the limit of {\em ideal} adiabatic 
conditions. We have also computed the corrections arising out of the
inevitable departures from ideal adiabaticity i.e the time of measurement being
large but finite. To overcome the {\em conceptual} difficulties with a
{\it quantum apparatus}, we have simulated a {\it classical apparatus} as
a {\em large} assembly of spin-1/2 systems.
We end this article with 
a conclusion and a discussion of some future issues.
\end{abstract}
\vskip 2pc]
\section{Introduction}
{\label{intro}}
Interpretation of measurements in quantum theory and the issue of reality
of wave functions have occupied a central position in physics. The conventional
wisdom is that the property of linear superposition of physical states
necessitates an ensemble interpretation of physical states(wave functions)
and the outcome of a measurement on a single quantum state is random and
consequently no reality, at least in the same sense of the word as was
associated in pre-quantum era, can be ascribed to a wave function. For generic
quantum states this is indeed the accepted view point and has been beautifully
supported even by recent developments like the no-cloning theorem etc. But
Aharonov and Vaidman \cite{av}, and subsequently Aharonov, Anandan and Vaidman 
\cite{aav} showed
that {\it adiabatic measurements} on a restricted class of {\em unknown single
quantum states} can preserve the original state in the limit of {\em infinitely
long measurement times} and at the same time give as output the {\em expectation
value} of an arbitrary observable. Since the original state is unaltered, it
can be used in further measurements of this type, and with the right number
of independent measurements, the state can be fully determined without 
affecting it. As shown by these authors the scheme works only when the original
unknown state is restricted to be a {\em nondegenerate eigenstate} of the system
Hamiltonian. The main caveat is that in every realisable set up the measurement
time can be made quite large but is never infinite. This leads to a correction
which is harmless for ensembles but critical for a single quantum state. As
shown in \cite{qureshihd} this correction precludes ascribing reality to
single wavefunctions of even this restricted class. However, the present
work may point to a way out of this difficulty.

In their original analysis \cite{av,aav} what played the role of apparatus was a
quantum mechanical system whose Hilbert space dimensionality is infinite. This
leads to a variety of technical issues which have been addressed in detail in \cite{qureshihd}. In this paper we have found an alternate model for adiabatic
(protective) measurements where we have replaced the 'apparatus' also by a
spin-1/2 system. To avoid various conceptual pitfalls of a 'quantum apparatus'
we have subsequently considered an extension whereby a large number of spin-1/2 particles
is made to simulate a 'classical' apparatus.

\section{Conventional measurement}
To set the stage for a discussion of the main contents of this paper, let
us first discuss the idea of a conventional (quantum)measurement. 

The current viewpoint, developed by the founding fathers of Quantum Mechanics,
is that the measurement of 
an observable corresponding to an operator $A$  of a system in the state
$\psi$ will have as its  outcome only the eigenvalues of $A$. The choice of
the eigenvalue is completely random and the system
after measurement will 'collapse' into the eigenfunction which corresponds
to the eigenvalue which is the outcome. As a consequence,the original state 
is irretrievably altered.Even though the individual outcomes are random,
the frequency with which the eigenvalue $\lambda_i$ occurs is given by
$|\langle \psi|\chi_i\rangle|^2$ where $|\chi_i\rangle$ is the corresponding
eigenstate.

Now we discuss how conventional or impulsive measurement is realised when
the apparatus is decsribed by an infinite dimensional Hilbert space.
Let $Q_S$ be an operator, corresponding to the observable of the system we 
wish to measure, and let it interact with the appropriate apparatus through 
an interaction,
\begin{equation}
H_I=\hbar g(t)Q_A\otimes Q_S
\end{equation}
where $Q_A$ is an observable of the apparatus,and $g(t)$ is the strength of 
the interaction normalized such that $\int_{0}^{\tau} g(t)=1$. The interaction
is nonzero only in the short interval $[0,\tau]$. Let the system be initially
in state $|\nu\rangle$ which is not necessarily an eigen state of $Q_S$, and the 
apparatus be in state $|\phi(r_0)\rangle$ which is a wave packet of eigen
state of the operator $R_A$ conjugate to $Q_A$, centered at the eigen 
value $r_0$. The interaction $H_I$ is of short duration, and assumed to be so 
strong that the effect of the free Hamiltonians of the apparatus and the system
can be neglected. More explicitly, the total Hamiltonian is given by 
\begin{equation}
H(t)=I_S\otimes H_A+H_S\otimes I_A +\hbar g(t)Q_A\otimes Q_S
\end{equation}
If $|t=0\rangle$ is the state vector of the combined apparatus-system just before
the measurement process begins, the state vector after T is given by,
\begin{equation}
\label{timeorder2}
|t=T\rangle={\cal T}e^{\frac{-i}{\hbar}\int_{0}^{T}H(\tau)d\tau}|t=0\rangle
\end{equation}
where ${\cal T}$ is the time odering operator.Consequently, after time $\tau$
the state of the system is
\begin{equation}
\label{tauorder}
|t=\tau\rangle=e^{\frac{-i}{\hbar}(I_S\otimes H_A+H_S\otimes I_A)\tau -iQ_A\otimes Q_S} |t=0\rangle
\end{equation}
By making $\tau$ suitably small, the terms depending on $H_A,H_S$ can be
neglecteds. Then the combined wave function of the system and the 
apparatus at the end of the interaction can be written as,
\begin{equation}
|\psi(\tau)\rangle=e^{-iQ_A\otimes Q_S}|\nu\rangle|\phi(r_0)\rangle
\end{equation}      
If we expand $|\nu\rangle$ in the eigenstates of $Q_S$, $|s_i\rangle$,we get,
\begin{equation}
|\psi(\tau)\rangle=\sum_ie^{-iQ_As_i}c_i|s_i\rangle|\phi(r_0)\rangle
\end{equation}
where $s_i$ are the eigen values of $Q_S$ and $c_i$ are the expansion coefficients.
The exponential term shifts the center of the wave packet by $s_i$:
\begin{equation}
\label{postimpulsive}
|\psi(\tau)\rangle=\sum_ic_i|s_i\rangle|\phi(r_0+s_i)\rangle
\end{equation}
To see this, first consider a {\it r-representation} of $|\phi(r_0)\ra$:
\br
\label{rrep}
|\phi(r_0)\ra &=& \int dr f(r-r_0)|R=r\ra\nonumber\\
&=& \int dq dr f(r-r_0)\la q|r\ra |q\ra
\er
On using $\la q|r \ra = e^{-iqr}$
\br
\label{displace}
e^{-iQ_As_i} &=& \int dq~dr f(r-r_0) \la q|r \ra e^{-iqs_i}|a\ra\nonumber\\
&=&dq~dr f(r-r_0) \la q| r+s_i\ra |q\ra\nonumber\\
&=& |\phi(r_0+s_i\ra
\er
In the last step we have shifted the variable of integration to $r+s_i$ and
used the r-representation eqn(\ref{rrep}) again.

The state given in eqn(\ref{postimpulsive}) is an entangled state, where the position of the wave packet gets correlated
with the eigen state $|s_i\rangle$. Detecting the center of the wave packet at 
$r_0+s_i$ will throw the system into the eigenstate $|s_i\rangle$. Thus we see
that impulsive measurement leads to complete destruction of the wave function
and once a measurement is made no further measurement can be made. Also the 
result of the outcome has to be statistically determined. Such a measurement
is realized through the Stern-Gerlach experiment where the apparatus is really 
the position of the silver atom and hence is a continious variable and the
Hilbert space is unbounded. 

Now we propose an apparatus to measure spin for a two-state  system(spin $1/2$) 
like the Stern-Gerlach apparatus but unlike the Stern-Gerlach apparatus
where the position is unbounded and is an observable operating on an infinite dimensional Hilbert space,
the Hilbert space of our {\it apparatus} will be taken to be finite dimensional. The main advantage
in using a finite dimensional Hilbert space is that our operators are also
bounded so perturbations can be applied safely because the interaction Hamiltonian 
will never be arbitarily large. Adiabatically measuring spin 
this way is a new concept which can be realized experimentally
with trapped atoms. The considerations of this paper can be potentially
significant for {\em reading out problem in quantum computation}.

We consider a measurement on a two-state system 
{\it S} (spin$1/2$) in which a quantum two-state system acts as a 
{\it detector} in the precise sense that information about the state of
the measured system is carried by the state of the {\it quantum detector}.
The Hilbert space $ {\cal H}_S$ of the system is spanned 
by the orthonormal states $|\uparrow\rangle$ and $|\downarrow\rangle$ 
while the states $|d_\uparrow\rangle$ and $|d_\downarrow\rangle$ 
span the detector ${\cal H}_D$ of the detector. The detector is 
initially in $|d_\downarrow\rangle$ state and clicks only when
the spin of the system is $|\uparrow\rangle$(\cite{zeh,wigner,scully,zurek}).
In other words, the {\em measurement interaction} can be represented as 
\begin{eqnarray}
\label{measint}
|\uparrow\rangle|d_{\downarrow}\rangle &~~{\stackrel{U_M}{\rightarrow}}& |\uparrow\rangle|d_{\uparrow}\rangle\nonumber\\
|\downarrow\rangle|d_{\downarrow}\rangle &~~{\stackrel{U_M}{\rightarrow}}& |\downarrow\rangle|d_{\downarrow}\rangle
\end{eqnarray}     
We try to figure out the most general unitary transformation $U_M$ that 
will lead to such an interaction: 
\begin{eqnarray}
U_M|\uparrow\rangle|d_{\downarrow}\rangle = |\uparrow\rangle|d_{\uparrow}\rangle e^{i\theta}\nonumber\\
U_M|\downarrow\rangle|d_{\downarrow}\rangle = |\downarrow\rangle|d_{\downarrow}\rangle e^{i\phi}
\end{eqnarray} 
The explicit form of $ U_M$ is 
\begin{equation}
\label{unitary}
U=
\left( \begin{array}{cccc}
0 & e^{i\theta} & 0 & 0 \\
{B_1}{e^{i\theta_1}} & 0 & {B_3}{e^{i\theta_3}}& 0 \\
\pm{B_3}{e^{i\phi_1}} & 0 & \pm{B_1}{e^{i\phi_3}} & 0 \\
0 & 0 & 0 & e^{i\phi}
\end{array} \right) 
\end{equation}
where $\theta$, $\phi$, $\theta_1$, $\theta_3$, $\phi_1$, $\phi_3$ are arbitary 
phases and can have any values and ${B_1}^2+{B_3}^2=1$(Due to the fact that $U$ 
is unitary). Thus there can be infinitely many such unitary transformations possible 
and hence infinitely many interaction Hamiltonians. We 
try to find a unitary transformation of the type in eqn(\ref{unitary}) which 
gives the simplest possible interaction Hamiltonian. 
The structure of the most general interaction Hamiltonian is given
by $H_I=\hbar g(t)\sum_{i,j}c_{ij}Q_{A}^{i}Q_{S}^{j}$ 
with $g(t)$ normalised according to  $\int_{0}^{\tau} g(t)=1$. For the 
{\em impulsive} case the interaction takes place during a 
short time interval $[0,\tau]$. 
$Q_{A}^{i}$ are observables corresponding to 
the apparatus and $Q_{S}^{i}$ observables corresponding to the 
system. The choice $\theta=0$, $B_1=1$, $B_3=0$, $\phi=0$, $\theta_1=0$, 
$\theta_3=0$, $\phi_1=0$, $\phi_3=0$ and taking only the positive signs 
one indeed gets a simple form for $U_M$ which also gives a simple form
of $H_I$( not always obvious): 
\begin{equation}
\label{unit}
U=
\left( \begin{array}{cccc}
0 & 1 & 0 & 0 \\         
1 & 0 & 0 & 0 \\                                    
0 & 0 & 1 & 0 \\                                 
0 & 0 & 0 & 1  
\end{array} \right) 
\end{equation}
leading to,
\begin{equation}
\ln U=
\left( \begin{array}{cccc}
\frac{i\pi}{2} & \frac{-i\pi}{2} & 0 & 0 \\
\frac{-i\pi}{2} & \frac{i\pi}{2} & 0 & 0 \\
0 & 0 & 0 & 0\\
0 & 0 & 0 & 0
\end{array} \right)
\end{equation}
The Hamiltonian is given by,
\begin{equation}
H_I=i\hbar g(t)\ln U.
\end{equation} 
From now onwards we adopt the convention $\hbar = 1$. The Hamiltonian
that leads to (\ref{unit}) is,
\begin{equation}
H_I=-\pi g(t)(P_{z,+}^S\otimes P_{x,-}^A)
\end{equation} 
where $P_{x,-}^A$ is the projection operator for the apparatus spin down 
along x direction and
$P_{z,+}^S$ is the projection operator for system spin up along z direction. 
More explicitly
\be
P_{z,+}^S=\frac{1+\sigma_z}{2} ~~~ P_{x,-}^A=\frac{1-\sigma_x}{2}
\ee
Thus $Q_A=P_{x,-}^A$, $Q_S= -\pi P_{z,+}^S$ in this case.

\section{Alternate model for protective measurements }
Aharonov and Vaidman \cite{av}, and subsequently Aharonov, Anandan and Vaidman \cite{aav} proposed a radically different approach to quantum
measurements where there is {\it no collapse of the wavefunction after 
measurement} 
to one of the eigenstates of the measured observable and remarkably {\it 
the original state is preserved} even after the
measurement. However this scheme has certain limitations; first of all the
scheme applies only when the original state of the system is a {\em
non-degenerate eigenstate} of its Hamiltonian. 
More importantly their claim is valid only in the limit of {\em ideal adiabaticity} i.e in the limit of the measurement time being infinite. They thought that 
their type of measurement would enable ascribing  {\em reality} to the 
wavefunction  
of a {\em single unknown} quantum state, albeit of a restricted class, 
departing from the conventional ensemble 
interpretation of quantum mechanics. This premise rested on the property of
their measurements which maintained the original unknown state while giving
full information about it. However, as shown by Qureshi and Hari Dass\cite{qureshihd} 
even the very small correction term arising out of the departure from ideal
adiabaticity can spoil this property in the case of measurements on a 
{\em single}i quantum state. Question of ascribing reality to the wavefunction
arises only in that case. In the case of ensembles, this tiny correction is
irrelevant, but so is the question of the reality of wavefunctions. 
Nevertheless this scheme is attractive
in a practical sense since a protective measurement on an ensemble of states
(with the restrictions mentioned above) maintains
the ensemble in its original state to a very high degree\cite{cold}

First we briefly recapitulate the idea of protective measurements. In contrast
to the case of impulsive measurements discussed earlier, the 
interaction
of the system with the apparatus now is {\it weak} and {\it adiabatic}. Hence
one cannot neglect the free Hamiltonians. As mentioned before,
the system is assumed to be in a non-degenerate eigenstate of its Hamiltonian.
Let the Hamiltonian of the combined system be
\begin{equation}
\label{instant}
H(t)=H_A\otimes I_S+I_A\otimes H_S+g(t)Q_A\otimes Q_S
\end{equation}
Where $H_A$ and $H_S$ are the Hamiltonians of the apparatus and the system. The
coupling g(t) acts for a long time $T$ and goes to zero smoothly before and 
after the interaction. It is normalized again by $\int_{0}^{T} g(t)=1$. 
Therefore 
$g(t)\approx\frac{1}{T}$ is small and constant for the most part. Corrections
due to $g(t)$ deviating from a constant have been carefully analysed in
\cite{cold} and shown to be negligible.
If $|t=0\rangle$ is the state vector of the combined apparatus-system just before
the measurement process begins, the state vector after T is given by,
\begin{equation}
\label{timeorder}
|t=T\rangle={\cal T}e^{-i\int_{0}^{T}H(\tau)d\tau}|t=0\rangle
\end{equation}
where ${\cal T}$ is the time odering operator. We divide the interval $[0,T]$ 
into
$N$ equal intervals $\Delta T$, so that $\Delta T=\frac{T}{N}$, and because the 
full Hamiltonian commutes with itself at different times during $[0,T]$
(since we have taken $g(t)=\frac{1}{T}$), we can write eqn (\ref {timeorder})as
\begin{equation}  
|t=T\rangle=\{e^{-i\Delta T(H_A+H_S+\frac{1}{T}+Q_AQ_S)}\}^N|t=0\rangle
\end{equation}
Let us now examine the case when $Q_A$ commutes with $H_A$. 
Let $|a_i\rangle$ be the simultaneous eigenstates of $Q_A$ and
$H_A$ such that $Q_A|a_i\rangle=a_i|a_i\rangle$ and $H_A|a_i\rangle=E_i^a|a_i
\rangle$. The $|a_i\ra$ form an orthonormal basis in the apparatus Hilbert
space ${\cal H}_A$. Let $|\chi_j\ra$ be any orthonormal basis spanning the
system Hilbert space ${\cal H}_S$. Let the {\em exact} eigenstates of the 
instantaneous Hamiltonian of eqn(\ref{instant}) be expanded as
\be
\label{expn}
|\psi_\mu\ra = \sum_{ij} c^\mu_{ij}|\chi_j\ra|a_i\ra
\ee
satisfying
\be
\label{schr}
\{H_A\otimes I_S+I_A\otimes H_S +{\frac{1}{T}}Q_A\otimes Q_S\}|\psi_\mu\ra = \lambda_\mu|\psi_\mu\ra
\ee
Substituting the expansion eqn(\ref{expn}) in eqn(\ref{schr}) one gets
\be
\sum_{kj} c^\mu_{kj}\{E_i^a+H_S+{\frac{a_i}{T}}Q_S\}|\chi_j\ra|a_k\ra=
\lambda^\mu\sum_{pj}c^\mu_{pj}|\chi_j\ra|a_p\ra
\ee
Taking the inner product of both sides of this equation with $|a_i\ra$ one
gets
\be
(E_i^a+H_S+{\frac{a_i}{T}}Q_S)\sum_j c^\mu_{ij}|\chi_j\ra = \lambda^\mu\sum_j c^\mu_{ij}|\chi_j\ra
\ee
On introducing $|{\bar\mu}_i\ra = \sum_j c^\mu_{ij}|\chi_j\ra$, this can
be recast as
\be
(E_i^a+H_S+{\frac{a_i}{T}}Q_S)|{\bar\mu}_i\ra = \lambda^\mu|{\bar\mu}_i\ra
\ee
So the exact eigenstates can be written in a factorized form 
$|a_i\rangle|\bar\mu_i\rangle$ where
$|\bar\mu_i\rangle$ are system states which depend on the eigenvalue of $Q_A$, 
i.e., they are the eigenstates of $\frac{1}{T}a_iQ_S+H_S$. In the
extreme adiabatic limit $T \rightarrow\infty$, the states $|{\bar\mu}_i\ra$
approach $|\mu\rangle$ which are
eigenstates of $H_S$ with eigenvalue $\mu$. 
Now let us assume the initial state of the system and apparatus to be a direct 
product of a non-degenerate eigenstate $|\nu\ra$ of $H_S$
and a wave-packet state of the apparatus $|\phi(r_0)\rangle$ that is centred
around $r=r_0$:
\begin{equation} 
|t=0\rangle=|\nu\rangle|\phi(r_0)\rangle
\end{equation}
Introducing the complete set of exact eigenstates in the above equation, 
the wave
function at time T can now be written as,
\begin{equation}
\label{pro}
|t=T\rangle=\sum_{i,\mu}e^{-iE(a_i,\mu)T}~
\langle {\bar\mu}_i|\nu\rangle\langle a_i|\phi(r_0)\rangle
|a_i\rangle|{\bar\mu}_i\rangle
\end{equation}
The exact instantaneous eigenvalues $E(a_i,\mu)$ can be written as
\begin{equation}
\label{exact}
E(a_i,\mu)=E_{i}^{a}+\frac{\langle{\bar\mu}_i|Q_S|{\bar\mu}_i\rangle a_i}{T}
+\langle{\bar\mu}_i|H_S|{\bar\mu}_i\rangle
\end{equation}
On using first order perturbation theory, 
\begin{equation}
\label{per}
|{\bar\mu}_i\rangle=|\mu\rangle+\frac{1}{T}\sum_{\chi\neq \mu}\frac{|\chi\rangle a_i(Q_S)_{\chi\mu}}{E_\mu-E_\chi}+O(\frac{1}{T^2})
\end{equation}
Using equation (\ref {per}), and the fact that $|\mu\rangle$ and $|\chi\rangle$ 
are 
orthogonal we get, to first order in ${\frac{1}{T}}$ 
\begin{equation}
\label{perenergy}
E(a_i,\mu)=E_i^a+\mu+\frac{a_i}{T}\langle\mu|Q_S|\mu\rangle+O(\frac{1}{T^2}) 
\end{equation} 
To obtain the leading order result, we need to keep $O({\frac{1}{T}})$ terms
in $E(a_i,\mu)$ but only the leading order terms for $|{\bar\mu}_i\ra$( this
is because $E(a_i,\mu)$ is multiplied by T in the exponent). 
Summing over $\mu$
in eqn(\ref{pro}) 
(as a result of which only the term with $\mu=\nu$ survives) 
we get,
\begin{equation}
\label{final}
|t=T\rangle\approx \sum_i e^{-i\nu T}
|\nu\rangle e^{-iE_i^aT-ia_i\langle Q_S\rangle_\nu}|a_i\ra\la a_i
|\phi(r_0)\rangle
\end{equation}
On using the identity
\be
\label{ident}
\sum_i e^{-iE_i^aT-ia_i\la Q_S \ra_\nu}|a_i\ra\la a_i| = 
e^{-iH_AT-i\la Q_S \ra_\nu Q_A}
\ee
we get
\be
\label{fin}
|t=T\rangle \approx e^{-i\nu T} |\nu\rangle e^{-iH_AT-i\la Q_S \ra_\nu Q_A}
|\phi(r_0)\rangle
\ee
In the case of systems governed by a Heisenberg algebra for which displacement
operators exist, we can further simplify this to
\be
|t=T\ra \approx 
e^{-i\nu T} 
e^{-iH_AT}
|\nu\rangle 
|\phi(r_0+\la Q_S \ra_\nu)\rangle
\ee
where we have made use of eqn(\ref{displace}) in the last step.
This is the central result of \cite{av,aav} and one sees that in the 
extreme adiabatic limit the apparatus measures the {\em expectation value}
of a system observable directly and that too without {\em disturbing}
the state of the system!
\subsection{The alternate model}
Now we consider the case when ${\cal H}_A$ represents a {\em finite 
dimensional} Hilbert space. To be specific, we take it to be a spin-1/2
system. Because of the finite dimensionality of the Hilbert space, the
algebra of observables can not be of the Heisenberg type and consequently
there are no {\em displacement operators}. However, the relevant algebra
now is the Lie algebra and we have rotation operators instead.

For $H_A$ we choose a {\it rotationally invariant} operator which in the
present case means it is a constant. Here too we have an improvement over
the original situation of \cite{aav}. Let this constant be $E^a$ so that
$E^a_i$ is independent of $i$. For the system Hamiltonian $H_S$ we take
the Hamiltonian of the spin-1/2 system in the presence of an {\em unknown}(to
the person making the measurements) magnetic field 
$ \vec B = B_0 \vec{\tilde n}$. Thus $H_S = - B_0 \vec\sigma\cdot\vec{\tilde n}$. Let the eigenstates of $H_S$ be $|\tilde\pm\ra$ which are obviously
{\em nondegenerate}. Furthermore,
$\vec\sigma .\vec{\tilde n}$$|\tilde\pm\rangle=\pm|\tilde\pm\rangle$.
Following the discussions of sec.II we choose the interaction Hamiltonian to be
$H_I=-{\frac{\pi}{T}}(P_{z,+}^S\otimes P_{x,-}^A)$.
For this case $Q_A$ and $H_A$ commute.

Let the initial unknown state of the system be $|\tilde +\ra$ and let its
representation in terms of the basis vectors of the system used in eqn(\ref{measint}) be     
\be
|\tilde+\rangle= \alpha|\uparrow\rangle + \beta|\downarrow\rangle 
\ee
which is an eigen function of ${\it H_S}$ and let the detector be initially 
in the state ${\it |d_{\downarrow}\rangle}$.

Putting together all of these details in eqn.(\ref {final}) we get, for our
model,
\begin{equation}
|t=T\rangle\approx e^{i B_0T-iE^aT}e^{i\pi \la P_{z,+}^S \ra_\nu P_{x,-}^A}|\nu\rangle|d_\downarrow\rangle
\end{equation}
Now $\langle P_{z,+}^S\rangle_\nu=|\alpha|^2$ and $P^A_{x,-}=\frac{I}{2}-{S^A_x}$
Hence the state after the measurement is,
\begin{equation}   
|t=T\rangle\approx 
e^{i( B_0 -  E^a)T+\frac{i\pi|\alpha|^2}{2} }
|\nu\rangle e^{{-iS^A_x\pi|\alpha|^2}}|d_\downarrow\rangle 
\end{equation}    
Thus we see that in our version of the protective measurements, the apparatus state instead of being displaced by an amount proportional to the expectation
value of the system observable being measured is actually {\it rotated} around
the x-axis by an angle that is proportional to the same expectation value.
Again, the original wavefunction is 
completely preserved in the limit of ideal adiabatic conditions. 
As the original state is preserved even after the measurement as in the treatments of \cite{av,aav} it can be used to 
determine all the expectation values $\langle S^S_x\rangle$,$\langle 
S^S_y\rangle$,
$\langle S^S_z\rangle $ thereby fully determining
the unknown initial state. 
\section{Corrections due to departure from ideal adiabaticity}
So far the treatments were based on ideal adiabatic conditions and neglected
the $\frac{1}{T}$ correction terms arising out of the measurement 
time $T$ being finite even though arbitarily large in actual experiments. 
There are essentially two distinct types of $\frac{1}{T}$ corrections that
arise. First arises out of $O(\frac{1}{T^2})$ corrections to eqn(\ref{perenergy}) contributing to corrections to {\em phases}. The second arises out of keeping
the $O(\frac{1}{T})$ corrections to $|{\bar\mu}_i\ra$ in eqn(\ref{per}). 
We give below the ingredients for calculating both types of corrections:
\begin{equation}
\label{one}
\begin{array}{rr}
|\bar\mu\rangle\langle\bar\mu|\nu\rangle\approx\delta_{\mu\nu}|\mu\rangle+\frac{1}{T}\sum_{k\neq\mu}\frac{|\mu\rangle\delta_{k\nu}a_i(Q_S)_{k\mu}^{\ast}}{E_\mu-E_k}\\
+\frac{1}{T}\sum_{k\neq\mu}\frac{|k\rangle\delta_{\mu\nu}a_i(Q_S)_{k\mu}}{E_\mu-E_k}
\end{array}
\end{equation}
\begin{equation}
\label{two}
\langle\bar\mu|Q_s|\bar\mu\rangle\approx\langle Q_S\rangle_\mu+\frac{2a_i}{T}\sum_{k\neq\mu}\frac{(Q_S)_{\mu k}(Q_S)_{k\mu}}{E_\mu-E_k}
\end{equation}
\begin{equation}
\label{three}
\langle\bar\mu|H_S|\bar\mu\rangle\approx\langle H_S\rangle_\mu+\frac{a_{i}^{2}}{T^2}\sum_{k\neq\mu}\frac{\langle H_S\rangle_k(Q_S)_{\mu k}(Q_S)_{k\mu}}{(E_\mu-E_k)^2}
\end{equation}
Using eqns(\ref{two,three}) it is easy to show that the corrections to $E(a_i,\mu)$ are:
\begin{eqnarray}
\label{four}
E(a_i,\mu)&\approx&E_{i}^{a}+\langle H_S\rangle_\mu+\frac{a_i}{T}\langle Q_S\rangle_\mu\nonumber\\
&+&\frac{a_{i}^{2}}{T^2}\sum_{k\neq\mu}\frac{\langle H_S\rangle_k(Q_S)_{\mu k}(Q_S)_{k\mu}}{(E_\mu-E_k)^2}
\end{eqnarray}
As discussed at length in \cite{qureshihd} the $O(\frac{1}{T})$ corrections to 
the post-measurement state are the ones that come in the way of a {\em truly
protective} measurements on {\em unknown single} quantum states. As far as
measurements on {\em ensemble of states} are concerned none of these correctionsare serious. Again the $O(\frac{1}{T})$ corrections to {\em phases} do not
alter the leading order conclusions about the preservation of the original
state, so we shall neglect such corrections to phases.

With $|\tilde +\ra = \alpha|\uparrow\ra + \beta|\downarrow\ra$ as the initial 
state $|\nu\ra$ of the system, the state $|\chi\ra$ orthogonal to it is $|\tilde -\ra = \beta^*
|\uparrow\ra - \alpha^*|\downarrow\ra $. 
Then we have:
$E_\nu=- B_0$, $E_\chi= B_0$, 
$\langle Q_S\rangle_\nu=-\pi|\alpha|^2$,  
$\langle Q_S\rangle_\chi=-\pi|\beta|^2$ and  
$(Q_S)_{ \chi\nu}=-\pi\alpha\beta$.
Consequently With $|\nu\ra = |\tilde +\ra $ we
separate the contributions from $ \mu = \tilde +,\tilde - $ in eqn(\ref{pro}):
\be
e^{-i(E^a -  B_0)T+ \pi|\alpha|^2 a_i}[|\tilde +\ra -{\frac {a_i\alpha\beta}
{2 B_0 T}} |\tilde -\ra]|a_i\ra\la a_i|d_{\downarrow}\ra
\ee
\be
{\frac {a_i\alpha^*\beta^*}{2 B_0 T}}e^{-i(E^a+ B_0)T +
\pi|\beta|^2a_i}|\tilde -\ra |a_i\ra\la a_i|d_{\downarrow}\ra
\ee
Using eqn(\ref{ident}) along with the definition $\eta = {\frac{\alpha\beta}{2B_0T}}$ we can combine these into the final result:
\br
& &|t=T\ra = e^{i B_0T+\pi|\alpha|^2P^A_{x,-}}|\tilde +\ra|d_{\downarrow}\ra +\nonumber\\ 
& &[\eta^* e^{-i B_0T+\pi|\beta|^2
P^A_{x,-}}-\eta e^{i B_0T+\pi|\alpha|^2
P^A_{x,-}}]P^A_{x,-}|\tilde -\ra|d_{\downarrow}\ra
\er
Using the properties of the projection operator $P^A_{x,-}$ we can simplify 
this to 
\br
& &|t=T\ra = e^{i B_0T+{\pi|\alpha|^2\over 2}-\pi|\alpha|^2S^A_x}|\tilde +\ra|d_{\downarrow}\ra +\nonumber\\ 
& &[\eta^* e^{-i B_0T+\pi|\beta|^2
}-\eta e^{i B_0T+\pi|\alpha|^2
}]|\tilde -\ra|d_{\downarrow}\ra_x
\er
We thus see that correcting the final expression to the order of $\frac{1}{T}$
there exists a finite probability of $O(\frac{1}{T^2})$ that the system is
transfered to another orthogonal,non-degenerate state of its Hamiltonian thus
leading to a complete loss of the original state. Hence we cannot carry out
protective measurement on a single state safely because there exists a low 
but finite probability of the system loosing its original state after 
measurement. Hence to safely carry it out we must use an ensemble of state 
rather than a single state so that we may identify which outcome corresponds 
to the case where the original state is being preserved. But then the whole
issue of the reality of wavefunctions becomes irrelevant. However, as pointed
out in \cite{cold}, the protective measurements are still useful because
they maintain the {\em purity} of the original ensemble to a very high degree
(of the order of $1-{\frac{c}{T^2}}$).
\section{Classical apparatus in terms of quantum spins}
Till now we have used a quantum mechanical system with Hilbert space ${\cal H}_A$ to describe the 'apparatus'. This is of course unsatisfactory as it leads to
unsurmountable technical and philosophical problems. In this section we make a
proposal as to how one may overcome this difficulty. The idea is to 
{\it simulate} a classical detector out of 
several quantum detectors. A large collection of microscopic quantum 
systems is expected to behave like a macroscopic system  and hence 
simlate a classical detector. Specifically, fluctuations in certain 
``macroscopic'' observables will be small. We consider a system which 
has ${\it N}$ quantum detectors described by the tensor product Hilbert
space ${\cal H}_1\otimes{\cal H}_2....\otimes {\cal H}_N$ initially in
the state $|d_{1\downarrow}\ra|d_{2\downarrow}\ra..|d_{N\downarrow}\rangle$.
It is easy to generalise the considerations of our paper till now to this
enlarged situation.
\begin{equation}
Q_A=-\pi\sum_{l=1}^N(I\otimes I\otimes... \otimes P^{(l)}_{x,-}\otimes .. 
I\otimes I)
\end{equation}
Putting $P_{x,-}^{(l)}=\frac{(I-\sigma_x^{(l)})}{2}$,we get
\begin{equation}
Q_A=\frac{-\pi I^A}{2}+\pi\sum_{l=1}I\otimes I\otimes.. S_x^{(l)}\otimes .. 
I\otimes I
\end{equation}
When this detector is used to detect the system state $|\tilde+\rangle$
using the interaction Hamiltonian $H_I=\frac{1}{T}(P_{z,+}^S\otimes Q_A)$, 
we get, in the extreme adiabatic limit,
\begin{eqnarray}
|t=T\rangle&\approx& e^{{i B_0T}+\frac{i\pi|\alpha|^2}{2}}\nonumber\\
& &(|\tilde+\rangle \Pi_{n=1}^{N}exp({-i\pi|\alpha|^2)S_{x,n}}|d_{n\downarrow}\rangle_z
\end{eqnarray}
Thus we see that each individual quantum spin making up the apparatus is 
rotated about the x-axis by the same 
magnitude. The relevant {\em macroscopic observable} for the apparatus is
the {\it total spin} $\vec S = \sum_{i=1}^N {\vec s}_i$.
The fluctuation in the mean of the total spin is
proportional to $\frac{1}{\sqrt N}$. Thus increasing $N$ reduces the 
fluctuations and make the detector becomes {\em essentially classical}.
\section{Conclusion and future issues}
In this article we have considered an alternate realisation of the idea of
protective measurements first discussed by \cite{av,aav}. The apparatus has
been replaced by a finite dimensional Hilbert space system. Apart from
overcoming the difficulties arising out of unbounded operators(discussed
in detail in \cite{qureshihd}), the algebra of observables changes from
the Heisenberg type to the Lie algebra of {\em rotation operators}. Instead
of the apparatus being {\em displaced} by an amount proportional to
the {\em expectation values} of measured observables as in the original
treatments \cite{av,aav}, in our case the apparatus spin is {\em rotated}
by an amount proportional to the respective expectation values. Unlike
in the original treatment the apparatus position corresponding to the case
where protective measurement fails because of corrections due to {\em finite
but large times of observations} is {\it predetermined}. The consequences
of this for protective measurements will be dealt in future publications.
We have also tried to overcome the conceptual issues of working with a
{\it quantum detector} by considering an {\em ensemble of detectors}. We
believe this is a potentially interesting way of addressing the thorny issues
of {\it Quantum Measurements}. This can also open up interesting ways for
{\it readouts from Quantum Computers}. This will also be explored in a future
publication.
\section{Acknowledgement}
AD would like to thank Institute of Mathematical Sciences for granting the
Summer Research fellowship for carrying out this research. AD would also like
to thank his family and friends for their encuoragement and support.

\end{document}